\documentclass[twocolumn]{aastex631}

\usepackage{chngpage}
\usepackage{booktabs}
\usepackage{amsmath}
\usepackage{xcolor}

\newcommand{\uvec}[1]{\boldsymbol{\hat{\textbf{\textit{#1}}}}}

\newcommand\RGMX{\bgroup\markoverwith{\textcolor{cyan}{\rule[0.5ex]{4pt}{1pt}}}\ULon}

\newcommand\ACCX{\bgroup\markoverwith{\textcolor{red}{\rule[0.5ex]{4pt}{1pt}}}\ULon}
\shorttitle{Mutually misaligned circumbinary planets}
\shortauthors{Childs et al.}
\graphicspath{{./}{figures/}}

\begin{document}

\title{Coplanar circumbinary planets can be unstable to large tilt oscillations in the presence of an inner polar planet}

\author[0000-0002-9343-8612]{Anna C. Childs}
\affiliation{Center for Interdisciplinary Exploration and Research in Astrophysics (CIERA) and Department of Physics and Astronomy Northwestern University,1800 Sherman Ave, Evanston, IL 60201 USA}
\author[0000-0003-2401-7168]{Rebecca G. Martin}
\affiliation{Nevada Center for Astrophysics, University of Nevada, Las Vegas, NV 89154, USA}
\affiliation{Department of Physics and Astronomy, University of Nevada, Las Vegas, 4505 South Maryland Parkway,
Las Vegas, NV 89154, USA}
\author[0000-0003-2270-1310]{Stephen Lepp}
\affiliation{Nevada Center for Astrophysics, University of Nevada, Las Vegas, NV 89154, USA}
\affiliation{Department of Physics and Astronomy, University of Nevada, Las Vegas, 4505 South Maryland Parkway,
Las Vegas, NV 89154, USA}
\author[0000-0002-4636-7348]{Stephen H. Lubow}
\affiliation{Space Telescope Science Institute, 3700 San Martin Drive, Baltimore, MD 21218, USA}
\author{Aaron M. Geller}
\affiliation{Center for Interdisciplinary Exploration and Research in Astrophysics (CIERA) and Department of Physics and Astronomy Northwestern University,1800 Sherman Ave, Evanston, IL 60201 USA}



\begin{abstract}
Mutually misaligned circumbinary planets  may form in a warped or broken gas disc or from later planet-planet interactions. With numerical simulations and analytic estimates we explore the dynamics of two circumbinary planets with a large mutual inclination.  A coplanar inner planet causes prograde apsidal precession of the binary and the stationary inclination for the outer planet is higher for larger outer planet orbital radius. In this case a coplanar outer planet always remains coplanar. On the other hand, a polar inner planet causes retrograde apsidal precession of the binary orbit and the stationary inclination is smaller for larger outer planet orbital radius. For a range of outer planet semi-major axes, an initially coplanar orbit is librating meaning that the outer planet undergoes large tilt oscillations.  Circumbinary planets that are highly inclined to the binary are difficult to detect -- it is unlikely for a planet to have an inclination below the transit detection limit in the presence of a polar inner planet.  These results suggest that there could be a population of circumbinary planets that are undergoing large tilt oscillations.
\end{abstract}


\section{Introduction}
While binary stars are ubiquitous in our galaxy and over 5,000 exoplanets have been found, only 41 circumbinary planets (CBPs) have been observed and confirmed thus far \citep{ps}.  The majority of these circumbinary planets are gas giants although a super-Earth as small as about $2 \, M_{\oplus}$ has been observed \citep{Orosz2019} and are all in orbits that are nearly coplanar to the binary.  The observed coplanarity is undoubtedly due to the difficulty of observing planets on highly misaligned orbits by current techniques \citep{Schneider, MartinD2014, Martin_2017,Zhang2019,Martin2021}. 
Based on considerations of disc evolution, if it is sufficiently fast, then we expect planets to be either coplanar or polar with respect to the binary orbit \citep[e.g.][]{Martin17}. While polar planets have not yet been found, there are several examples of polar discs \citep{Kennedy2012,Kennedy2019,Kenworthy2022}. 

Polar circumbinary discs are more likely to be found around more eccentric binaries.  Misaligned planets may form more easily around wider binaries where binary eccentricities are higher and the disc evolution is slower \citep[e.g.][]{Czekala2019}.
We show here, that under certain circumstances, circumbinary planets can reside on orbits that undergo large tilt oscillations. In effect, a coplanar planet can undergo tilt oscillations from coplanar to beyond polar, almost retrograde.


Around an eccentric binary, there are two types of nodal precession of a misaligned circumbinary test particle \citep{Verrier2009,Farago2010,Doolin2011,Naoz2016}. An initially low inclination particle nodally precesses about the binary angular momentum vector (this is a circulating orbit). For high initial inclination, the particle can precess about the binary eccentricity vector (this is a librating orbit).  The polar stationary inclination is the inclination where a particle does not undergo nodal precession and remains highly inclined. The stationary inclination for polar orbits is at $90^\circ$ for all test particle semi-major axes. Prograde apsidal precession of the binary can be driven by general relativity \citep[e.g.][]{Naoz2017,Zanardi2018} or a triple star \citep[e.g.][]{Innanen1997,Morais2012}, and this leads to an increase in the polar stationary inclination with particle semi-major axis \citep{Lepp2022,Lepp2022b}.   With apsidal precession of the binary, beyond a critical semi-major axis, all particle orbits are circulating and the particle behaves as it would around a circular orbit binary.

A gaseous circumbinary disc that is in good radial communication can undergo similar nodal precession to a test particle \citep[e.g.][]{Papaloizou1995,Larwoodetal1996,Aly15}.   Differential nodal precession as a function of distance in such a disc results in viscous dissipation.  Viscous dissipation in a misaligned disc leads to evolution towards coplanar alignment or a stable polar configuration \citep{Martin17, Martin18,Lubow2018,Zanazzi2018,Cuello2019}.  Such misaligned circumbinary gas discs are often observed in nature and may result from the turbulent collapse of the molecular disc or from other mechanisms which later misalign the disc \citep{Offner2010, Tokuda2014, Bate2012, Bates2010, Bate2018, Bonnell1992,Nealon2020}. 

If a disc is not in good radial communication, the torque from the binary can lead to disc warping or breaking \citep[e.g.][]{Nixon2013,Facchini2013}. As a consequence of the breaking, an inner ring may then align (to polar or coplanar depending on its initial inclination and the binary eccentricity) on a shorter timescale than the outer parts of the disc \citep[e.g.][]{Lubow2018,Smallwood2020}.  The radius at which the disc breaks depends on the disc properties (such as the aspect ratio and viscosity) and the binary properties, but for standard parameters the disc can break close to the binary (at radii less that about ten times the binary semi-major axis) in both the viscous and wave-like regimes \citep{Facchini2013, Nixon2013, Lubow2018}.  Another mechanism that may result in CBP misalignment involves multiple accretion events onto a binary that can form misaligned discs \cite{Bate2018}.

Theoretical studies suggest that planet formation in polar circumbinary discs can take similar pathways as planet formation in coplanar circumbinary discs \citep{Childs2021coplanar,Childs2021polar,Childs2022}.  Giant planets that form in a warped or broken disc may form with a mutual misalignment. The misalignment could also arise from later planet-planet or planet-binary interactions \citep{Chen2022}.

In this Letter we investigate the dynamics of a two planet circumbinary system in which the planets have a mutual misalignment.  Understanding the dynamics of such a system will aid in future observations.  Polar planets present additional challenges for detection than coplanar planets. However, unique detectable dynamical signatures of coplanar planets that result from interactions with inner polar planets may provide indirect detections of polar planets. In Section~\ref{sec:Simulations} we present numerical simulations of the four-body system. 
We show that there is a range of semi-major axes for which an initially coplanar outer planet is librating and undergoing large tilt oscillations.  In Section~\ref{sec:analytic_estimates} we provide an analytic framework to find the stationary inclination of an outer planet with an inner circumbinary planetary companion and the range of semi-major axes for which the outer planet librates. We show good agreement with our numerical simulations.  Finally, in Section~\ref{sec:conclusions} we conclude with a summary of our findings.

\section{Numerical simulations}\label{sec:Simulations}

To model the dynamics of the four-body system we use the $n$-body code \textsc{rebound} \citep{Rein2012} with the \textsc{whfast} integrator \citep{Rein2015}.  We include the effects of GR by using the ``gr\_full" module from \textsc{reboundX} \citep{Tamayo2020}\footnote{ The $n$-body simulation results can be reproduced with the {\sc rebound} code (Astrophysics Source Code Library identifier {\tt ascl.net/1110.016}) and the {\sc reboundX} code (Astrophysics Source Code Library identifier {\tt ascl.net/2011.020}).}, but we find their effects to be small.  The binary is composed of two equal mass stars, $m_1=m_2=0.5 \, M_{\odot}$ with a total mass of $m_{\rm b}=m_1+m_2=1 \, M_{\odot}$. They are in an orbit with semi-major axis $a_{\rm b}= 0.5 \, \rm au$ with an eccentricity of $e_{\rm b}=0.8$.  We vary the mass, $m_{\rm p1}$, and initial inclination, $i_{\rm p1}$,  of the inner planet but keep it at a fixed semi-major axis of $a_{\rm p1}=5 \, a_{\rm b}$.  
For the outer planet, we vary the semi-major axis, $a_{\rm p2}$, and inclination, $i_{\rm p2}$, but fix the mass to a small value of $m_{\rm p2}=1\times 10^{-10} \, M_{\oplus}$. Such a small mass does not affect the dynamics of the four-body system.  Moreover, there is little difference in the behaviour if we instead use a Jupiter mass outer planet or a different binary mass fraction \citep[see also][]{Chen2019,Martin2019}. Both planets are in initially circular orbits.
\begin{figure}
	\includegraphics[width=\columnwidth]{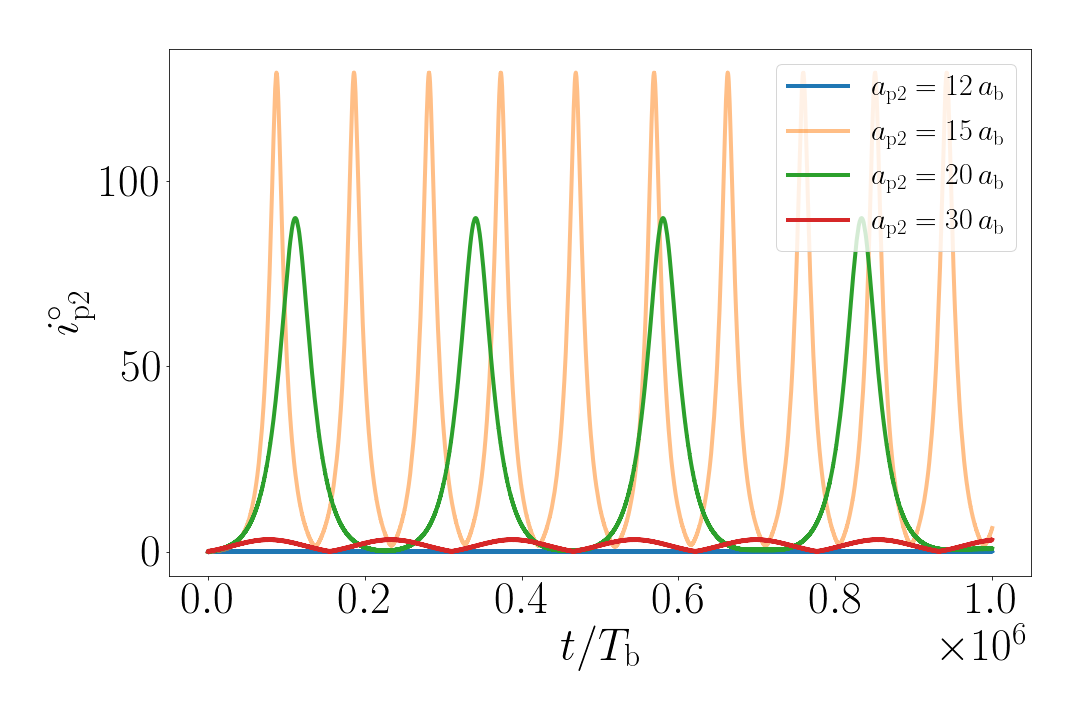}
    \caption{Numerical results for the inclination evolution of an outer planet orbiting an equal mass binary star with binary eccentricity $e_{\rm b}=0.8$ with  an inner Jupiter-mass polar planet orbiting with semi-major axis $a_{\rm p1}=5\,a_{\rm b}$.   The outer planet begins coplanar to the binary orbit with $i_{\rm p2}=0^{\circ}$ for different outer planet semi-major axis.}
    \label{fig:inc_evol}
\end{figure}

We measure all orbital elements in the frame of the binary. The inclination of a body is defined as 
\begin{equation}
    i =\textrm{cos}^{-1} (\uvec{l}_{\rm b} \cdot \uvec{l}_{\rm p})
\end{equation}
where $\, \uvec{} \,$ denotes a unit vector, $\textbf{\textit{l}}_{\rm b}$ is the binary angular momentum vector, and $\textbf{\textit{l}}_{\rm p}$ is the planet angular momentum. The nodal phase angle of a planet is calculated with
\begin{equation}
    \phi = \textrm{tan}^{-1} \left( \frac{\uvec{l}_{\rm p}\cdot (\uvec{l}_{\rm b} \times \uvec{e}_{\rm b})}{\uvec{l}_{\rm p} \cdot \uvec{e}_{\rm b}} \right ) + \frac{\pi}{2},
\end{equation}
where $\uvec{e}_{\rm b}$ is the eccentricity vector of the binary \citep{Chen2019}. Initially we take $\phi_{\rm p1}=\phi_{\rm p2}=90^\circ$.

\begin{figure*}

	\includegraphics[width=2\columnwidth]{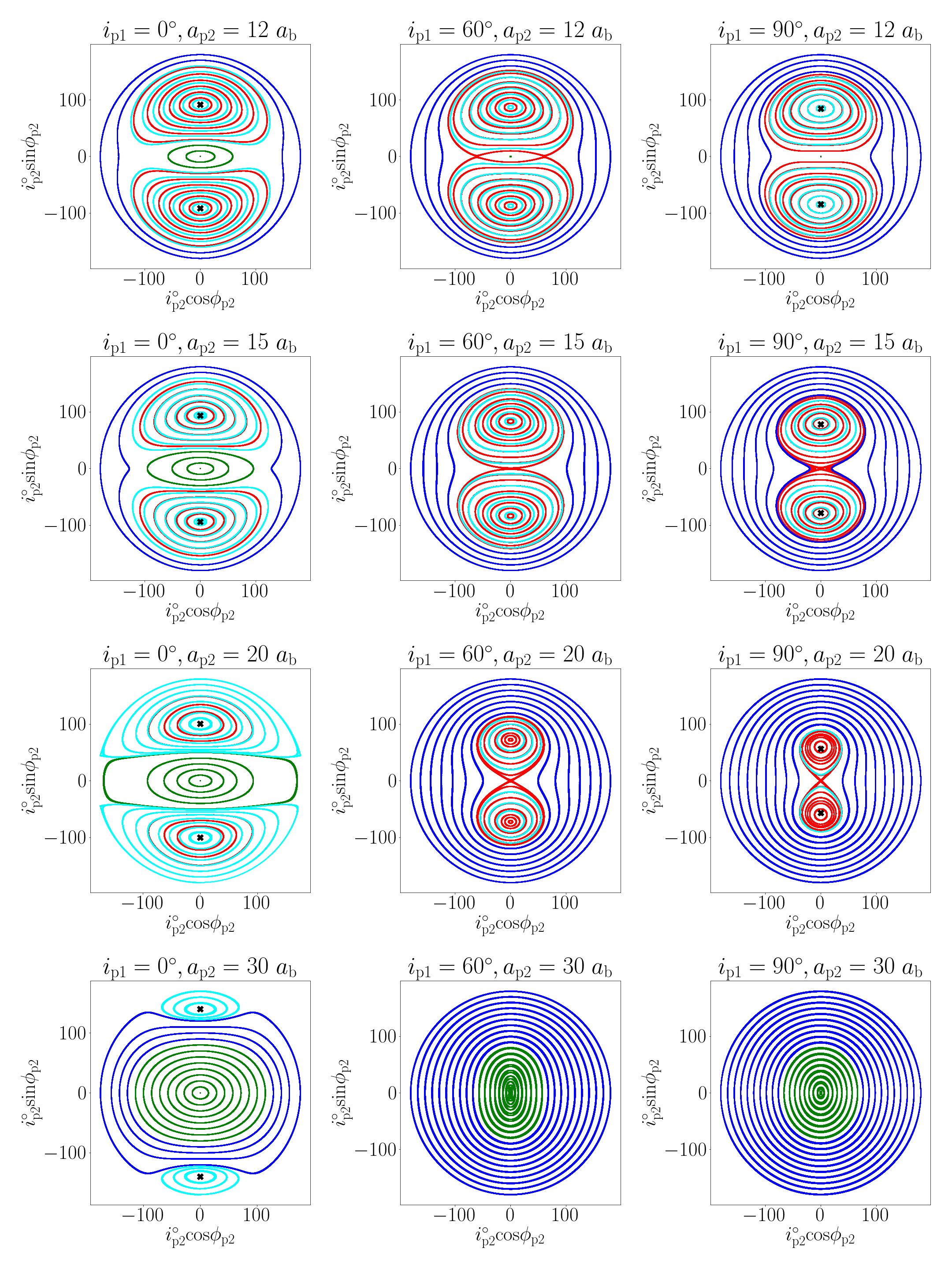}
    \caption{Phase diagrams for various semi-major axes of the outer planet and inclinations of the inner planet.   Green and blue orbits are circulating orbits with initial inclinations less than and greater than $90^{\circ}$, respectively.  Red and cyan orbits are librating orbits with initial inclinations less than and greater than the stationary polar inclination, respectively. The black crosses show the stationary polar inclination found with equation~(\ref{polar}).}
    \label{fig:phase_diagrams}
\end{figure*}

We first follow the inclination evolution of an outer planet with an inner polar planet companion that has the mass of Jupiter.  Figure~\ref{fig:inc_evol} shows the inclination of the outer planet over one million binary orbits ($T_{\rm b}$). The planet has an initial inclination of $i_{\rm p2}=0^{\circ}$ and we simulate the system at four different semi-major axes for the outer planet.
With $a_{\rm p2}=12 \, a_{\rm b}$, the planet remains nearly coplanar, although its inclination immediately jumps to about $ 0.1^{\circ}$ and  then slightly oscillates about this inclination.  This inclination is above the transit detection limit and, even for this small tilt, the planet spends about $94 \, \%$ of its time above the transit detection limit for an eclipsing binary \citep{Li2016}.   When the planet is farther out at $15 \, a_{\rm b}$ and $20 \, a_{\rm b}$, it undergoes large tilt oscillations.  In these cases,  after the initial evolution, the planet inclination never drops sufficiently for the planet to be detectable by periodic transits. For even larger outer planet semi-major axis, $a_{\rm p2}=30 \, a_{\rm b}$, the tilt oscillations are smaller. However, such a planet never has a small enough inclination to be below the transit detection limit even though it remains close to coplanar.

 Figure \ref{fig:phase_diagrams} shows  $i_{\rm p2}\cos \phi_{\rm p2}-i_{\rm p2}\sin \phi_{\rm p2}$ phase diagrams for various semi-major axes of the outer planet and inclinations of the inner planet. The inner planet has a fixed mass of $m_{\rm p1}=1 \, M_{\rm Jup}$ for each phase diagram.  The green and blue lines show circulating orbits with initial inclinations less than and greater than $90^{\circ}$, respectively.  The red and cyan lines show librating orbits with initial inclinations less than and greater than $90^{\circ}$, respectively. 

For an inner planet that is coplanar (left panels of  Figure \ref{fig:phase_diagrams}), a coplanar outer planet always remains coplanar. There are always circulating orbits for initial inclinations of the outer planet that are close to coplanar. For larger values of the outer planet semi-major axis, the stationary polar inclination is larger and so is the range of initial inclinations for circulating orbits. Outside of a critical radius, the librating orbits disappear and the only possible orbits are circulating. This is a result of the prograde precession of the binary that is driven by the inner planet. This is similar to the case of GR driven apsidal precession \citep{Lepp2022} except that the timescale for the precession may be much shorter in this case and therefore the critical radius is smaller.


For a polar inner planet (right panels  of Figure \ref{fig:phase_diagrams}), the binary undergoes retrograde nodal precession. The stationary polar inclination is smaller for larger semi-major axis of the outer planet. The critical inclination that separates the circulating and librating orbits also is smaller and the possibility for a coplanar orbit is removed for sufficiently large semi-major axis. This can be seen in the two middle panels on the right hand side where an initially coplanar orbit is librating! The initially coplanar outer planet undergoes large tilt oscillations. For even larger semi-major axis, all possible orbits become circulating. Similar behaviour can also be seen in the middle panels for an inner planet inclination of $i_{\rm p1}=60^\circ$.

\begin{figure}
	\includegraphics[width=\columnwidth]{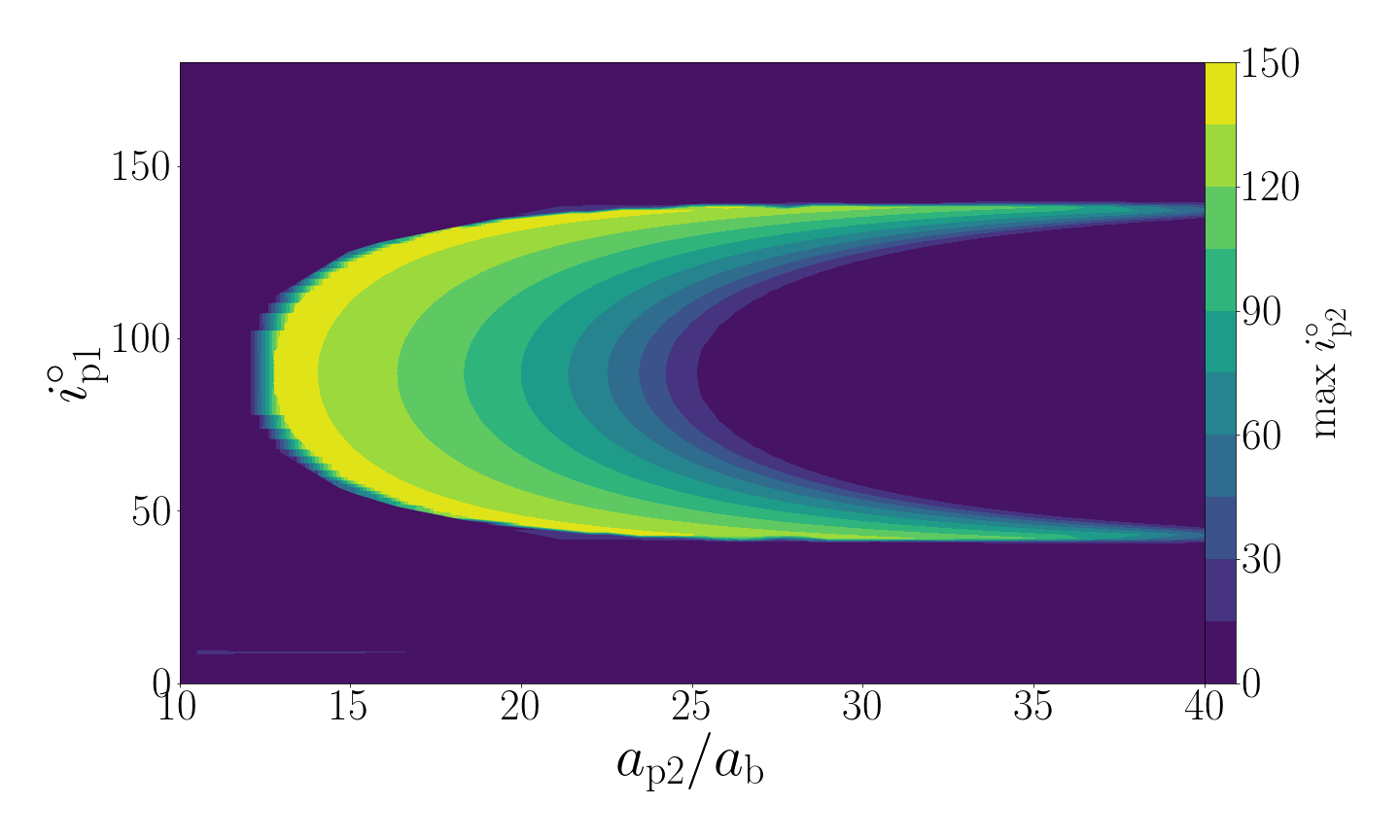}
    \caption{Contour plot of the maximum inclination of the outer planet as a function of the outer planet semi-major axis and the initial inclination of the inner Jupiter-mass planet. 
    The outer planet has an initial inclination of $0^{\circ}$.}
    \label{fig:contours}
\end{figure}

Figure \ref{fig:contours} shows how the inclination of a Jupiter mass inner planet affects the maximum inclination of the outer planet, which begins coplanar, over one million binary orbits as a function of the outer planet's distance. 
If the outer planet is closer in than $12 \,a_{\rm b}$ the outer planet remains coplanar regardless of the inner planet's inclination.  This is because the dynamics of the outer planet here are dominated by the dynamics of the binary, and the outer planet  precesses with the binary.  Exterior to $12 \, a_{\rm b}$ however, the outer planet can be librating, and therefore its maximum inclination can be very large. This occurs for  inner planet inclination in the approximate range of $40-130^\circ$.   For larger outer planet semi-major axis, the planet again remains coplanar since it no longer precesses with the binary. The distance at which this happens depends strongly on the inner planet inclination.  



\section{Analytic Estimates}\label{sec:analytic_estimates}

\begin{figure*}
	\includegraphics[width=2\columnwidth]{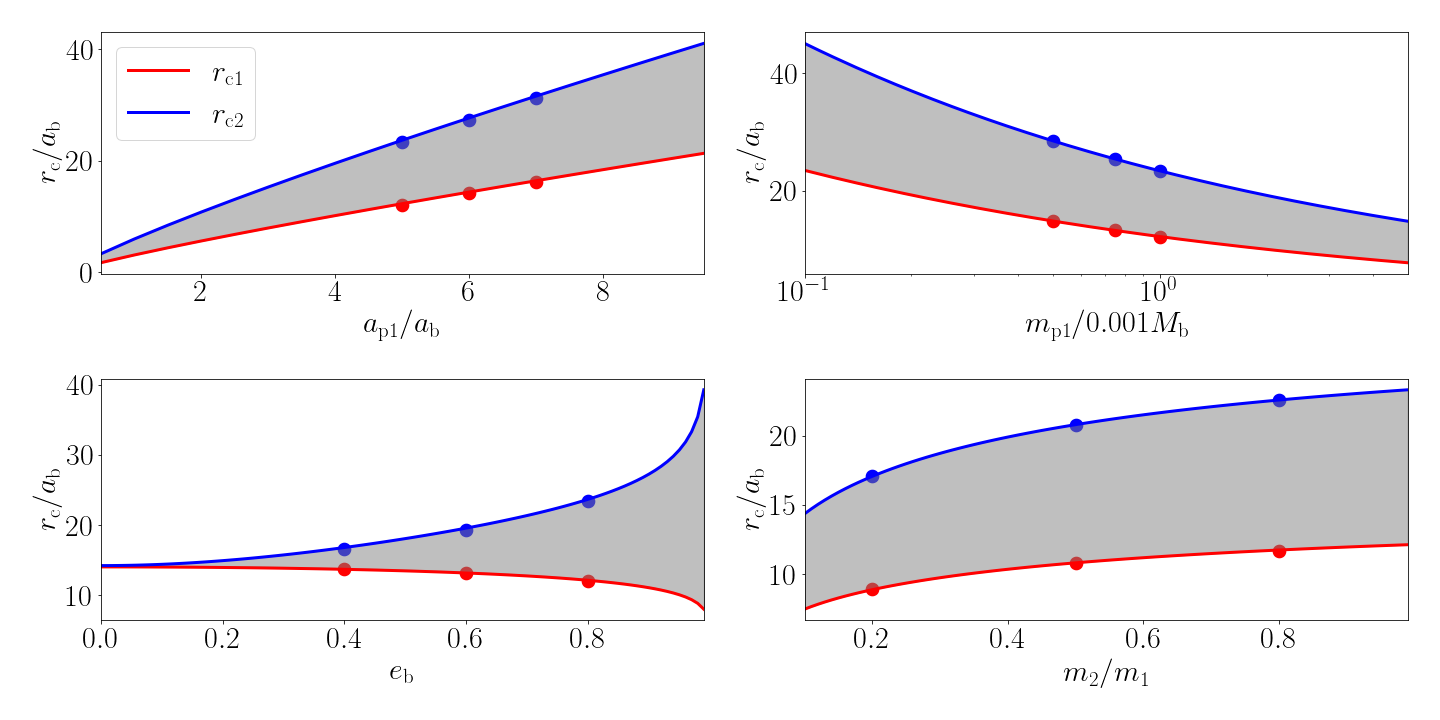}
    \caption{Critical radii as a function of $a_{\rm p1}$ (top left), $m_{\rm p1}$ (top right), $e_{\rm b}$ (lower left), and $m_2/m_1$ (lower right).  The default parameters from our numerical solutions are used for all other parameters besides the parameter we vary in each panel. The red and blue lines mark  $r_{\rm c1}$ and $r_{\rm c2}$  (equations~(\ref{eq:rc1}) and~(\ref{eq:rc2}), respectively).  The gray shaded region between these two curves denotes where an initially coplanar outer planet is librating with an inner polar planet.  The points show our numerical results.}
    \label{fig:rc_params}
\end{figure*}

We now use an analytic model to find the critical semi-major axes at which the outer planet orbital dynamics changes in the presence of an inner polar planet.  The outer planet is modelled as a test particle and both planets are in initially circular orbits. 
 We make the approximation that the gravitational effects of the inner planet on the outer planet are ignored. The inner planet interacts with the binary and causes it to precess. The test particle experiences the gravitational effects of the binary only. We examine the accuracy of this approximation below. In addition we consider only effects that arise in the quadrupole approximation for the gravitational forcing by the binary. The circumbinary planet orbit remains nearly circular because its eccentricity
is constant in time in the quadrupole approximation \citep{Farago2010}.
This quadrupole approximation for the binary is justified because octupole terms vanish 
for a circular orbit circumbinary planet \citep[e.g, equation (7) of][]{Elia2019}.
 
 The nodal precession of the outer planet is described by
\begin{equation}
    \dot \phi_{\rm p2}=\dot \phi_{\rm p2,binary}-\dot{\varpi}_{\rm b}.
\end{equation}
The time evolution of the nodal precession of the test particle orbiting around an eccentric binary up to the quadrupole level of secular  approximation is
\begin{equation}
\dot \phi_{\rm p2,binary}=-\frac{3}{4}\Omega_{\rm b} \frac{m_{1}m_{2}}{m_{\rm b}^2}\left(\frac{a_{\rm b}}{a_{\rm p2}}\right)^{7/2}
F_1,
\label{dphibdt}
\end{equation}
where the orbital frequency of the binary is $\Omega_{\rm b}=\sqrt{G m_{\rm b}/a_{\rm b}^3}$
and  we define
\begin{equation}
    F_1=\cos i_{\rm p2}\left[1+\frac{3}{2} e_{\rm b}^2-\frac{5}{2}e_{\rm b}^2 \cos 2\phi_{\rm p2}\right]
\end{equation}
\citep[e.g.][]{Innanen1997,Kiseleva1998,Naoz2017,Zanardi2018}. 

The binary undergoes apsidal precession driven by the inner planet. 
The precession rate  of the longitude of periastron of the binary in the limit of a small mass companion is
\begin{equation}
    \dot{\varpi}_{\rm b} = \frac{3}{4} \Omega_{\rm b} \frac{m_{\rm p1}}{m_{\rm b}}
    \left(\frac{a_{\rm b}}{a_{\rm p1}}\right)^{3}
    F_2,
\end{equation}
 where
\begin{align}
    F_2  =  (1-e_{\rm b}^2)^{-1/2}\left[ \right. & 
    2-2e_{\rm b}^2+5(e_{\rm b}^2-\sin^2i_{\rm p1})\sin^2 \omega_{\rm b} \cr
 &     -
    \cos i_{\rm p1}\left(1-e_{\rm b}^2+5e_{\rm b}^2\sin^2 \omega_{\rm b} \right)
   \left. \right]
\end{align}
\citep[e.g.][]{Innanen1997,Naoz2016} and the argument of periastron of the binary is $\omega_{\rm b}$. 

We find the polar stationary inclination of the outer planet (by setting $ \dot \phi_{\rm p2}=0$ and $\phi_{\rm p2}=90^\circ$) to be
\begin{equation}
    \cos i_{\rm s}=- \frac{m_{\rm p1}m_{\rm b}}{m_1m_2}
    \left(\frac{a_{\rm p2}}{a_{\rm b}}\right)^{7/2}  
    \left(\frac{a_{\rm b}}{a_{\rm p1}}\right)^{3} 
    \frac{F_2}{1+4e_{\rm b}^2}.
    \label{polar}
\end{equation}
For a coplanar inner planet, $i_{\rm p1}=0^\circ$, the value of $\omega_{\rm b}$ does not matter because those terms cancel out to give $F_2=(1-e_{\rm b}^2)^{1/2}$. For a polar inner planet, $i_{\rm p1}=90^\circ$, $\omega_{\rm b}=90^\circ$ and we find $F_2=-3(1-e_{\rm b}^2)^{1/2}$ \citep[see also][]{Zhang2019}. The black crosses in Fig.~\ref{fig:phase_diagrams} show that equation~(\ref{polar}) accurately predicts the stationary polar inclination for a two circumbinary system with $i_{\rm p1}=0^\circ$ and  $90^{\circ}$.  
For the $i_{\rm p1}=90^{\circ}$ system, for larger semi-major axis of the outer planet, the stationary inclination is slightly higher than the analytical predictions.  This small offset can be attributed to GR effects included in the simulations that are not accounted for in our analytic expressions and use of the quadrupole approximation of the potential in the analytic model.


Now we consider the critical $a_{\rm p2}$ where the librating orbits reach an inclination of $0^\circ$ (by setting $i_{\rm p2}=\phi_{\rm p2}=0^\circ$ and $\dot \phi_{\rm p2}=0$) to be
\begin{equation}\label{eq:rc1}
    r_{\rm c1}
     = \left[\left(\frac{a_{\rm p_1}}{a_{\rm b}}\right)^{3}
     \frac{m_1m_2}{m_{\rm p1}m_{\rm b}}
     \frac{(1-e_{\rm b}^2)}{(-F_2)}\right]^{2/7}a_{\rm b}.
\end{equation}
For the polar inner planet parameters, this is $r_{\rm c1}=12.1\,a_{\rm b}$, in agreement with the right hand panels of Fig.~\ref{fig:phase_diagrams} and Fig.~\ref{fig:contours}. There is also a critical radius outside of which there are no librating orbits ($i_{\rm p2}=0^\circ$, $\phi_{\rm p2}=90^\circ$ and $\dot \phi_{\rm p2}=0^\circ$) given by
\begin{equation}\label{eq:rc2}
    r_{\rm c2}
     = \left[\left(\frac{a_{\rm p_1}}{a_{\rm b}}\right)^{3}
     \frac{m_1m_2}{m_{\rm p1}m_{\rm b}}
     \frac{(1+4e_{\rm b}^2)}{(-F_2)}\right]^{2/7}a_{\rm b}.
\end{equation}
For the polar inner planet parameters we find $r_{\rm c2}=23.4\,a_{\rm b}$, in agreement with the right hand panels in Fig.~\ref{fig:phase_diagrams} and Fig.~\ref{fig:contours}.
Now if a coplanar outer planet has a radius in the range $r_{\rm c1}<a_{\rm p2}<r_{\rm c2}$, then it is in a librating orbit. 
A planet that is initially coplanar with the binary at a larger orbital radius that the outer critical radius $r_{c2}$  or a smaller radius than the inner critical radius $r_{c1}$ is on a circulating orbit about the binary, provided that is it not too close to the binary.



Figure \ref{fig:rc_params} shows how these critical radii change as a function of $a_{\rm p1}$, $m_{p1}$, $e_{\rm b}$, and $m_2/m_1$.  The default values from our numerical solutions are used for all other parameters in addition to the parameter we vary in each panel. The  parameters are $a_{\rm p1}=5\,a_{\rm b}$, $m_{\rm p1}=0.001\,m_{\rm b}$, $e_{\rm b}=0.8$ and $m_2/m_1=1$. The radial range for an outer librating planet that is initially coplanar increases significantly with binary eccentricity. The radius is larger for larger inner planet semi-major axis and smaller with the inner planet mass.  The critical radii are relatively insensitive to the binary mass fraction.  We numerically find the critical radii and plot these numerical values on Figure \ref{fig:rc_params} with points.  We find good agreement for $r_{\rm c1}$ and $r_{\rm c2}$ with our analytic predictions and the location of these points are insensitive to the effects of GR in the parameter space sampled. 


To ensure our predictions are applicable to circumbinary systems, we estimate the disk breaking radius using equation~(34) of \cite{Lubow2018}.  Assuming the fiducial disk parameters of \cite{Lubow2018}, a disk aspect ratio of $H/R=0.05$, and our binary parameters, the disk breaking radius is $r_{\rm break} \approx 2.8 \, a_{\rm b}$.  This is in agreement with simulations of similar disks \citep{Martin18,Abod2022}.  This disk breaking radius may permit the existence of an inner highly misaligned planet ($a_{\rm p1} \lesssim 2.8 \, a_{\rm b}$) and an outer planet beyond $r_{\rm c1}$ ($a_{\rm p_1} \gtrsim 12 \, a_{\rm b}$).

We consider the accuracy of our approximation that ignores the gravitational effects of the inner planet on the outer planet. The ratio $R$ of the nodal precession of the outer planet that is caused by the inner planet 
to the nodal  precession rate  of the outer planet that is caused by the binary is estimated by using Equation (\ref{dphibdt}) as 
\begin{equation}
R = \frac{m_{\rm p_1} m_{\rm b}}{m_1 m_2} \left(\frac{a_{\rm p_1}}{a_{\rm b}}\right)^{2} 
\frac{\tan{(i_{\rm p_2})}}{1+\frac{3}{2} e_{\rm b}^2},
\end{equation}
where we have adapted Equation (\ref{dphibdt})  for the inner planet to the case of a 'binary' consisting of the inner planet and the central binary as a point mass. 
Note that $R$ is independent of $a_{\rm p_2}$.
For the parameters adopted in this paper, $R \simeq 0.05 \tan{i_{\rm p_2}}$. Therefore, for most tilt angles of the outer planet, the nodal precession of the outer planet is dominated by the effects of the binary. We also made a direct numerical test of this approximation by running the model with $a_{p_2}=15a_{\rm b}$ in which the gravitational forcing of the outer planet due to the inner planet is ignored and compared the result to the case in which that forcing is included. We find that the difference in $i_{p2}(t)$ for these two cases is small, $<1\%$, which justifies our approximation.

\section{Conclusions}\label{sec:conclusions}

Using numerical and analytical methods we have explored the dynamics of two mutually misaligned circumbinary planets. We focused on a polar inner planet and an initially coplanar outer planet. Such a configuration could arise from disc evolution.  We treat the outer planet as a test particle and allow the inner planet to have a range of masses. Based on our tests, the properties of the outer particle well represent the properties of a planet.  We find that the inner planet  drives apsidal precession of the binary, at a faster rate than GR, which  affects the dynamics of the outer planet. 

A polar inner planet  causes retrograde apsidal precession of the binary orbit and the stationary inclination is smaller for larger outer planet semi-major axis.  There is a range of semi-major axis for the outer planet for which an  outer planet that is initially coplanar with the binary is  on a librating orbit and therefore undergoes large tilt oscillations (see Fig,~\ref{fig:inc_evol}). 
Outside this range of radii, a planet that is initially coplanar with the binary is on a circulating orbit.
The radial extent of the librating region increases with larger inner planet semi-major axis, binary eccentricity and mass ratio, and decreases with the mass of the inner planet.
With an inner polar planet, an outer planet that begins close to coplanar with the binary may spend only a small fraction of its time with an inclination small enough to be detected by periodic transits. Planets that are undergoing large tilt oscillations may never have a small enough inclination to be detected in this way.



We predict that there is a large radial region where initially nearly coplanar orbits to the binary undergo large tilt oscillations,  if there is an inner highly misaligned companion.  Transit detection techniques are strongly biased against finding such highly misaligned CBPs.  These results can help constrain occurrence rates of, and aid future observations of highly misaligned CBPs.

\begin{acknowledgements}
ACC and AMG acknowledge support from the NSF through grant NSF AST-2107738. RGM and SHL acknowledges support from NASA through grants 80NSSC19K0443 and 80NSSC21K0395. SHL thanks the Institute for Advanced Study for visitor support.
\end{acknowledgements}

\bibliography{main}{}
\bibliographystyle{aasjournal}

\end{document}